\documentclass[conference]{IEEEtran}
\IEEEoverridecommandlockouts

\usepackage{cite}
\usepackage{amsmath,amssymb,amsfonts}
\usepackage{algorithmic}
\usepackage{graphicx}
\usepackage{textcomp}
\usepackage{xcolor}
\usepackage{braket}
\def\BibTeX{{\rm B\kern-.05em{\sc i\kern-.025em b}\kern-.08em
    T\kern-.1667em\lower.7ex\hbox{E}\kern-.125emX}}
\begin{document}
\title{Demonstrating Record Fidelity \\  for the Quantum Fourier Transform}

\author{
\IEEEauthorblockN{Philipp Aumann\IEEEauthorrefmark{1}\IEEEauthorrefmark{4}\thanks{PA and MF contributed equally to this work and share first authorship.}, Michael Fellner\IEEEauthorrefmark{1}\IEEEauthorrefmark{5}, David Alber\IEEEauthorrefmark{1}, Max Cykiert\IEEEauthorrefmark{2}, Christoph Fleckenstein\IEEEauthorrefmark{1},\\ Roeland ter Hoeven\IEEEauthorrefmark{1}, Leo Stenzel\IEEEauthorrefmark{2}, Riccardo J. Valencia Tortora\IEEEauthorrefmark{1} and Wolfgang Lechner\IEEEauthorrefmark{1}\IEEEauthorrefmark{2}\IEEEauthorrefmark{3}}
\IEEEauthorblockA{\IEEEauthorrefmark{1} Parity Quantum Computing GmbH, A-6020 Innsbruck, Austria}
\IEEEauthorblockA{\IEEEauthorrefmark{2}Parity Quantum Computing Germany GmbH, 20095 Hamburg, Germany}
\IEEEauthorblockA{\IEEEauthorrefmark{3}Institute for Theoretical Physics, University of Innsbruck, A-6020 Innsbruck, Austria}
\IEEEauthorblockA{\IEEEauthorrefmark{4}p.aumann@parityqc.com}
\IEEEauthorblockA{\IEEEauthorrefmark{5}m.fellner@parityqc.com}
}

\maketitle

\begin{abstract}
    We demonstrate the Parity Architecture on quantum hardware, using the quantum Fourier transform (QFT) as a benchmark. As a result, a record performance in both fidelity and qubit count is achieved using quantum processors with a native CZ-based instruction set. On the IBM Heron r3 chip, a process fidelity of the QFT algorithm of ${F \approx 10^{-2}}$ for ${N=50}$ qubits is achieved. The scaling of the speedup compared to previous SWAP-based methods is super-exponential $\mathcal{O}(\exp(N^2))$. Furthermore, we show that the scaling can be improved further by including iSWAP gates in the instruction set. 
\end{abstract}

\section{Introduction}
While tremendous improvements in gate fidelities, qubit lifetimes and quantum hardware design have been made in the last few years~\cite{li2023, Gyger2024, Xue2024, Liang2025, Chiu2025, Valentini2025, Bland2025, marxer2025, Loschnauer2025, Hughes2025}, efficient implementations of quantum algorithms remain essential for obtaining the best performance from state-of-the-art quantum computers. Both algorithm and hardware development must go hand in hand to push these limits even further.

On the algorithmic side, in particular, reducing qubit counts, gate counts and circuit depth is crucial for both near term applications~\cite{Preskill2018, Cerezo2021, Bharti2022} and fault-tolerant quantum computing~\cite{Zhou2025, Ruiz2025, Gidney2025, Webster2026}. Quantum algorithms require a high degree of coupling between qubits, in contrast to the natural nearest neighbor interactions provided on many contemporary quantum devices. This mismatch between the problem- and hardware connectivity is commonly addressed using SWAP networks~\cite{Crooks2018, Ogorman2019} along linear qubit chains, which have been conjectured to be optimal for certain fully-connected circuit constructions, even on devices with denser connectivity graphs such as square grids~\cite{Weidenfeller2022}. This conjecture was recently disproved by the Parity Twine approach~\cite{Dreier2025}, which outperforms SWAP networks in both circuit depth and gate count.

Given a compilation strategy, a key step is identifying a suitable metric for assessing its performances. To this end, the quantum Fourier transform (QFT) represents a natural application-level benchmark for quantum processors~\cite{Lubinski2023} since it is a cornerstone subroutine for a plethora of applications~\cite{Kitaev1995, Shor1997, Hallgren2005, Harrow2009, Childs2010}, including Shor's period finding. The latter has been recently been proposed as one of the key performance indicators for quantum computing~\cite{Zimboras_2025}.
\begin{figure}[t!]
    \centering
    \includegraphics[width=1\columnwidth]{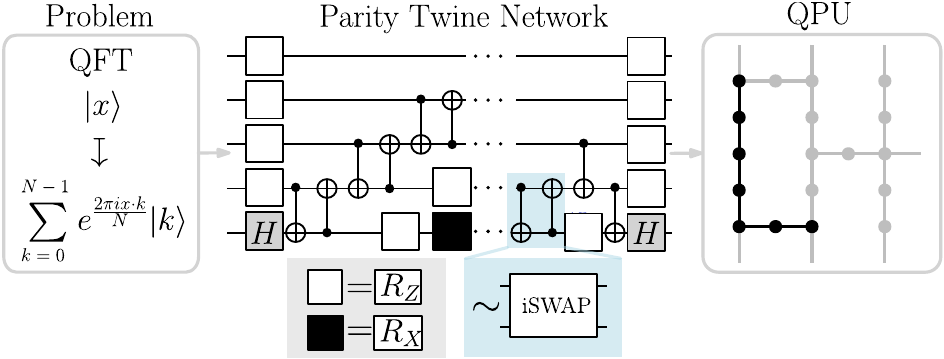}
    \caption{Overview of the workflow: The QFT implements the quantum analog to the discrete Fourier transform of an input state $\ket{x}$. It is realized by a quantum circuit consisting of single qubit rotations and a Parity Twine Network which efficiently creates correlations via chains of DCNOTs. A single DCNOT is locally equivalent to an iSWAP gate (blue box). This implementation respects the device qubit topology and does not require any qubit routing.}
    \label{fig:overview}
\end{figure}

In this work, we present the highest fidelity ever reported for the unitary QFT to date, which we achieved by using a combination of the efficient SWAP-less Parity Twine method~\cite{Dreier2025, Klaver2026}  and the \texttt{ibm\_boston} Heron r3 device, one of the most performant QPUs to date with a CNOT-like native gate set. Most importantly, we demonstrate a super-exponential scaling improvement compared to previous methods in experiments. Additionally, we explore the impact of different native gate sets on the algorithm performance, demonstrating the optimal scaling for the QFT by utilizing iSWAP gates in conjunction with the Parity Twine method.  

The remainder of this paper is organized as follows. We introduce the fundamentals of the Parity Twine technique in Sec.~\ref{sec:twine}, present and analyze our experimental results in Sec.~\ref{sec:experiments}, and conclude in Sec.~\ref{sec:conclusion}.

\section{Parity Twine}\label{sec:twine}
Parity Twine Networks (PTNs)~\cite{Dreier2025, Klaver2026} provide a systematic approach for compiling quantum algorithms that require a large number of pairwise two-qubit interactions, such as the QFT and the Quantum Approximate Optimization Algorithm (QAOA)~\cite{Farhi2014}. Building on the universal Parity Architecture~\cite{Lechner2015, Fellner2022a, Fellner2022b}, which enables efficient and fault-resilient quantum computation at the cost of a quadratic qubit overhead, PTNs emulate its key principles on arbitrary hardware connectivity graphs without increasing the qubit count. Their central goal is to minimize the two-qubit gate count and circuit depth under realistic hardware connectivity constraints.

This is achieved by reorganizing the computation around the \textit{parity information}, which encodes the relative alignment of logical variables. By operating on parity information rather than physically moving qubit states, PTNs avoid SWAP gates, which decompose into three two-qubit entangling gates each. Instead, parity information is moved along the physical qubits using sequences of CNOT gates, including double-CNOT (DCNOT) sequences, significantly reducing gate count and circuit depth. A DCNOT gate consists of two anti-parallel CNOT gates. The PTN used for the QFT experiments is depicted in Fig.~\ref{fig:overview}.

While PTNs were originally designed for architectures with CNOT-like~\cite{Krizan2025} entangling gates, providing highly efficient algorithm realizations, they can alternatively be designed to use iSWAP gates as the basic entangling gates. As iSWAP gates are locally equivalent to DCNOT gates, a DCNOT chain can be replaced by an iSWAP chain (see Fig.~\ref{fig:overview}), further halving the number of required two-qubit gates. In that case, when using a linear PTN, each required interaction can then be implemented using only one iSWAP gate on average, yielding the theoretical asymptotic lower bound for the two-qubit gate count for algorithms requiring all-to-all interactions such as the QFT. Therefore, the iSWAP-based PTN is optimal on a linear chain. As the DCNOT-based Parity Twine method on a linear chain already reduces the gate count by two thirds~\cite{Dreier2025} compared to the best previous SWAP-network based implementation strategies~\cite{Fowler2004, Maslov2007, Takahashi2007, Holmes2020, Park2023}, the iSWAP-based Parity Twine only requires one third of the gate count compared to the prior algorithms. 

\section{Experimental Results}\label{sec:experiments}
We execute the unitary QFT on state-of-the-art superconducting qubit processors from IBM and Rigetti, and benchmark its performance for the Parity Twine and the SWAP-network compilation strategies. 

The end-to-end performance of QFT is measured using the \textit{process fidelity}~\cite{Baeumer2024}. The protocol is based on preparing $m$ states $\ket{s_l}$ for which the QFT yields a single computational basis state ${\ket{k_l}\in \{\ket{j}, 0\leq j \leq 2^N-1\}}$, i.e., ${\text{QFT}\ket{s_l}=\ket{k_l}}$. These initial states can be prepared with high fidelity using solely single qubit rotations.
We then compute the process fidelity
\begin{align}\label{eq:process_fidelity}
    \mathcal{F}_{\mathrm{proc}} &\approx \frac{m}{m-1} \left[\frac{1}{m}\sum_{l=1}^m\sqrt{ \text{Pr}\left(k_l|  \tilde{\mathcal{QFT}}(\ket{s_l})\right)}\right]^2\\ \notag 
    &\quad - \frac{1}{m(m-1)}\sum_{l=1}^m \text{Pr}\left(k_l|  \tilde{\mathcal{QFT}}(\ket{s_l})\right),
\end{align}
for the QFT executed on hardware, denoted as $\tilde{\mathcal{QFT}}$, where ${\text{Pr}(x | \rho)}$ is the probability to measure the output bitstring $x$ given a density matrix $\rho$.

Unless otherwise stated, we have chosen ${m=20}$ initial states and $2000$ shots per circuit evaluation for the QFT experiments in this work, in accordance with Ref.~\cite{Baeumer2024}. 

\subsection{Demonstrating Record Fidelity}
\begin{figure}[t!]
    \centering
    \includegraphics[width=1\columnwidth]{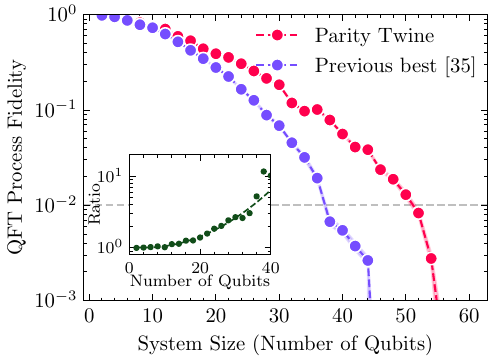}
    \caption{Results of QFT experiments as detailed in section~\ref{sec:experiments}, comparing the QFT process fidelity for different system sizes of the Parity Twine method against the best alternative state-of-the-art implementation introduced by Fowler et al.~\cite{Fowler2004}. The data is obtained from running both methods on the \texttt{ibm\_boston} Heron r3 device, using $XX$ dynamical decoupling and layouting by the \texttt{qiskit} transpiler.  The gray dashed line indicates the threshold of $10^{-2}$. The inset shows the ratio between the process fidelities, highlighting the super-exponential gain of Parity Twine over the best known alternative. The dashed line in the inset corresponds to the fit function ${1.0416e^{0.0015n^2-0.0134n}}$.} Note that the error-bars, given by the shaded areas, are hardly visible due their small size.
    \label{fig:boston_experiments}
\end{figure}

Figure~\ref{fig:boston_experiments} depicts the QFT process fidelity from Eq.~\eqref{eq:process_fidelity} for the Parity Twine method compared to the best previous implementation, introduced by Fowler et al.~\cite{Fowler2004}, evaluated on the \texttt{ibm\_boston} backend with a heavy-hexagonal qubit topology using $XX$ dynamical decoupling~\cite{PhysRevLett.82.2417}. Note that the techniques in Ref.~\cite{Fowler2004}, although established two decades ago, still represent the state-of-the art implementation of the QFT via SWAP-networks.

The physical qubit layouts were chosen by the \texttt{qiskit} transpiler with optimization level $3$ for every instance, in order to maximize the probability of using the best qubit set available. For the sake of completeness, we present the same benchmark on a fixed set of qubits in Appendix~\ref{appendix:fixed_layout}, which yields a significantly lower overall fidelity for all implementation strategies. 

\begin{figure}[t!]
    \centering
    \includegraphics[width=1
\columnwidth]{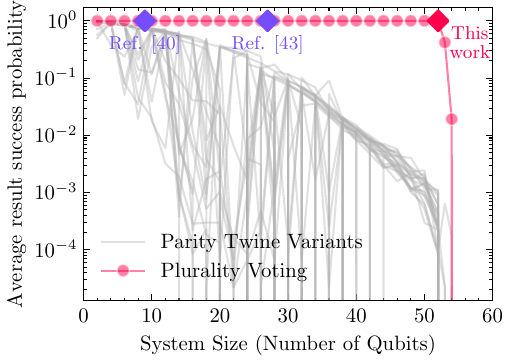}
    \caption{Post-processed results of the QFT experiments using plurality voting~\cite{maksymov2023}. For each system size and instance, we construct $21$ variants (gray data) of the linear Parity Twine method with $m=10$ initial states and $1000$ shots each. Additionally, we apply $XX$ dynamical decoupling and plurality voting with a threshold of ${t=7}$. Here, we compare against unitary implementations of the QFT, where Ref.~\cite{Baeumer2024} reported an average result success probability of close to 1 for up to 9 qubits and Ref.~\cite{Chen2024} for up to 27 qubits. In this work, we establish a new record by pushing this threshold to 52 qubits.}
    \label{fig:plurality_voting}
\end{figure}

We observe that the performance decreases significantly faster with the number of qubits $N$ in the best baseline implementation in comparison to the Parity Twine implementation. In particular, the Parity Twine QFT algorithm retains a process fidelity of $>10^{-2}$ for ${N\sim 50}$ qubits, in stark contrast to the best known alternative, which yields values above this threshold only up to ${N\sim 36}$ qubits. 
The sudden jumps between data points, e.g., ${N=32}$ and ${N=34}$ for Parity Twine and between ${N=36}$ and ${N=38}$ for the previous best method originate from a change in the qubit layout induced by the qiskit transpiler. 

The improved scaling of the process fidelity with $N$ for the Parity Twine method is directly linked to the reduced two-qubit gate count and circuit depth required, compared to the Fowler method~\cite{Dreier2025}. Specifically, since Parity Twine requires $\mathcal{O}(N^2)$ fewer two-qubit gates than the Fowler implementation, the expected improvement in the process fidelity is super-exponential, i.e., $\mathcal{O}(\exp(N^2))$, which we verify experimentally in this work. (see inset in Fig.~\ref{fig:boston_experiments}). This is further enhanced by a linear reduction in circuit depth, which leads to shorter idling time per qubit.

Although a heavy-hex-tailored PTN was introduced in Ref.~\cite{Dreier2025}, it does not outperform the linear PTN on \texttt{ibm\_boston}. This arises because the heavy-hex PTN reduces the leading-order two-qubit gate count of $n^2$ to ${5/6 n^2}$ compared to the linear PTN, but at the cost of an increased depth ($5n$ vs. $4n$), making the performance dependent on the relative impact of idling versus gate errors. We therefore report results obtained using the linear PTN only~\cite{Klaver2026}.

In a further step, we post-process the QFT data using the plurality voting algorithm introduced in Ref.~\cite{maksymov2023}. The results are shown in Fig.~\ref{fig:plurality_voting}. 
Plurality voting aggregates results of measurements in $N_\nu$ different, equivalent circuit variants.
It suppresses errors which only occur in at most $t$ variants, where $t$ is the threshold of the plurality voting,
but it also increases the outcome probability for the most-likely state.
Thus, its result should not be interpreted as a fidelity, and we refer to it as the \textit{average result success probability}.
Plurality voting is commonly used for post processing of QFT experiments \cite{maksymov2023,Chen2024,Baeumer2024}. For a more detailed discussion, we refer to Appendix~\ref{appendix:plurality_voting}.
Because plurality voting combines multiple circuit executions and applies classical post-processing, the resulting average success probability should not be compared directly with the process fidelity values. We therefore report plurality voting results separately and use them only as a benchmark of the maximum achievable success probability under this mitigation strategy.

The data required for the plurality vote are gathered by running $21$ different variants of the Parity Twine QFT implementation on the \texttt{ibm\_boston} backend including dynamical decoupling. 
Each variant samples $m=10$ initial states for each system size with $1000$ shots. With a threshold of ${t=7}$, we achieve an average result success probability of close to 1 for up to 52 qubits, not only surpassing the best previous result of 9 qubits achieved in Ref.~\cite{Baeumer2024} on the \texttt{ibm\_kyiv} backend, but also clearly exceeding the 27-qubit QFT previously reported on an ion-based quantum computer~\cite{Chen2024}, thereby establishing a new world record.
Even though the result quality obtained from plurality voting depends on the number of variants, the  threshold and the quantum hardware used, we show that the best reported result can be obtained by combining Parity Twine as the optimal implementation together with a state-of-the-art QPU.

We note that a higher average result success probability can be achieved using a non-unitary dynamic implementation of the QFT by employing measurements and classically-controlled corrections, as reported in Ref.~\cite{Baeumer2024}, which is useful if the QFT appears at the end of a quantum circuit. However, since our goal is to showcase Parity Twine as a method for synthesizing unitary circuits, we restrict our comparison to unitary implementations of the QFT.

\subsection{Improved scaling with iSWAP gates}
We also evaluate the performance of the iSWAP-based Parity Twine QFT. Figure~\ref{fig:ratios} directly compares the results obtained on the Rigetti \texttt{Ankaa-3} device (see Appendix~\ref{appendix_rigetti_experiment}), which offers a native iSWAP gate, and on the \texttt{ibm\_boston} (Heron r3) device (see Fig.~\ref{fig:boston_experiments}), where CZ is the native two-qubit entangling gate. Specifically, we plot the ratio of the two QFT process fidelities for the state-of-the-art QFT implementation~\cite{Fowler2004} as well as for our Parity Twine method on a linear chain.
We observe that, while the \texttt{ibm\_boston} processor consistently yields higher fidelities (i.e., all ratios are smaller than 1), the slope is significantly smaller for the Parity Twine approach. That indicates that the Parity Twine approach better exploits the native iSWAP gates, further boosting its performance. 

\begin{figure}
    \centering
    \includegraphics[width=\columnwidth]{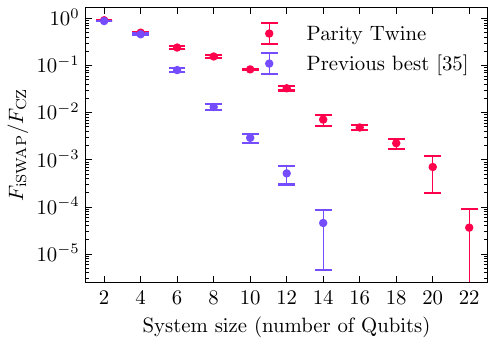}
    \caption{Ratio of the process fidelity for the QFT implementation on the CZ-based \texttt{ibm\_boston} device ($F_{\text{CZ}}$) and the Rigetti \texttt{Ankaa-3} hardware with native iSWAPs ($F_{\text{iSWAP}}$). The ratio is smaller than 1 in all cases, indicating that the IBM hardware is less susceptible to quantum errors. However, the smaller slope for the Parity Twine method indicates a prominent benefit in gate count for iSWAP-based devices with our approach.}
    \label{fig:ratios}
\end{figure}

\section{Discussion and Outlook}\label{sec:conclusion}
The results presented in this work demonstrate the largest-scale and highest-fidelity experimental implementation of a unitary quantum Fourier transform to date. Our experiments confirm that the reduction of gate count and circuit depth via Parity Twine significantly improves the process fidelity of the QFT over the best known alternative. 

Furthermore, on platforms supporting native iSWAP gates, Parity Twine circuits achieve optimal two-qubit gate count. By comparing process fidelities of the CZ-based device with the iSWAP-based on, we show that the Parity Twine approach fully exploits the native iSWAP gate over the best known alternative.

While this work experimentally focuses on the quantum Fourier transform, the underlying connectivity challenge also appears in other highly connected algorithms. A notable example is the Quantum Approximate Optimization Algorightm (QAOA), 
making the advantages of Parity Twine relevant in the broad field of quantum optimization~\cite{Dreier2025} as already shown in Ref.~\cite{MontanezBarrera2025}. Experimental validation across additional algorithms remains an important direction for future work. Beyond algorithm-specific improvements, Parity Twine holds promise as a general problem compilation technique, exhibiting enhanced performance---particularly on complex, highly-connected problems---compared to current state-of-the-art compilers~\cite{Dreier2025}. Efforts are underway to release this Parity Twine compiler for wider use.

In addition, while we focused on linear PTNs in this work, the Parity Twine method is broadly applicable across hardware platforms, including ladder-shaped and square-lattice qubit topologies~\cite{Dreier2025}. A detailed experimental study of more complex PTNs is left for future work.

Beyond near-term hardware, connectivity constraints also remain a central challenge for fault-tolerant quantum architectures. In surface-code-based approaches, logical qubits are typically arranged with limited connectivity, and long-range interactions require additional operations such as lattice-surgery procedures. While the present work focuses on demonstrating the hardware-level benefits of Parity Twine without introducing additional logical qubit overhead, these results motivate further investigation of Parity-based compilation strategies in fault-tolerant settings.

\appendices
\section{Experiments on a Heavy-Hexagonal QPU with fixed layouts (IBM Heron)}\label{appendix:fixed_layout}
\begin{figure}[ht!]
    \centering
    \includegraphics[width=\columnwidth]{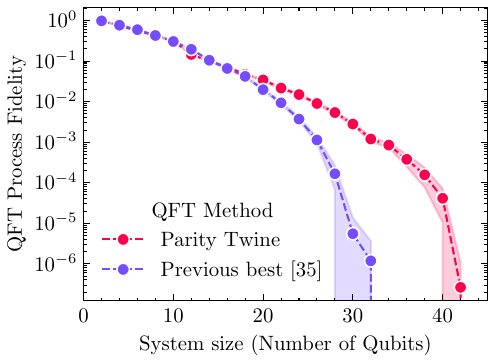}
    \caption{Results of QFT experiments on the \texttt{ibm\_boston} backend, where the physical qubits were fixed along a line on the device instead of transpiler-based layouting.}
    \label{fig:fixed_line}
\end{figure}

Our main results, shown in Fig.~\ref{fig:boston_experiments}, were obtained by running the \texttt{qiskit} transpiler on each instance with unrestricted choice of qubit layouts. To isolate the effect of layout selection, we repeated the QFT experiments using a fixed, pre-defined linear arrangement of physical qubits. The same set of physical qubits was used for each instance across both methods. As shown in Fig.~\ref{fig:fixed_line}, enforcing a fixed layout results in the expected overall decrease in gate fidelity, but also reduces fluctuations in the performance metric that were previously caused by variable layout choices. This data was acquired without any error-mitigation strategies. The overall trends in the data remain largely unchanged despite the constraint on qubit placement.

\section{Plurality Voting}\label{appendix:plurality_voting}
Plurality voting~\cite{maksymov2023} is an error-mitigation method which aggregates measurements from multiple circuit variants. Variants are different quantum circuits that would yield equivalent results in the absence of noise.
Plurality voting is commonly used for post processing measurements from QFT experiments~\cite{maksymov2023,Chen2024,Baeumer2024}.
The idea is as follows:
\begin{enumerate}
    \item Sample one measurement from each variant;
\item Determine the most-frequently observed state across all variants;
\item If this state appeared more than $t$ times, 
accept it and add it to the outcome distribution; 
\item  Repeat (1-3) until outcome distribution converges.
\end{enumerate}

Plurality voting helps to suppress systematic error in individual variants.
Being a non-linear aggregation, it also creates a skewed outcome distribution.
Thus, it should only be used for experiments where a single state is expected as ideal outcome.

Our implementation of plurality voting avoids the sampling process, 
and instead computes the probability for each state to be measured in more than $t$ variants simultaneously,
as described in Ref.~\cite{Chen2024}.

This approach is only exact if $t \geq N_v/2$ where $N_v$ is the number of variants.
Thus, we choose $t = \lceil N_v/2\rceil$, or the largest number of variants any state has been measured in,
whichever is smaller.
For smaller $t$, this effectively includes states which appear more than $t$ times, 
but are not the most likely ones.
To make this approximation consistent, 
we also count states multiple times if they appear in more than ${2t+1}$ variants.

\section{Experiments on an iSWAP QPU (Rigetti Ankaa-3)}\label{appendix_rigetti_experiment}
\begin{figure}[ht!]
    \centering
    \includegraphics[width=\columnwidth]{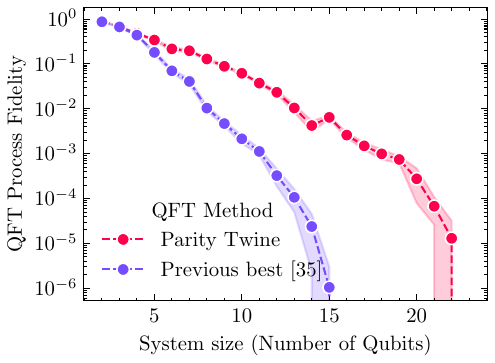}
    \caption{Results of QFT experiments on the Rigetti \texttt{Ankaa-3} backend using native iSWAP gates, comparing Parity Twine with the best alternative implementation.}
    \label{fig:rigetti_experiments}
\end{figure}

In Fig.~\ref{fig:rigetti_experiments} we show the QFT process fidelity on Rigetti's \texttt{Ankaa-3} QPU with native iSWAP gates for various system sizes using the Parity Twine implementation and the best previous method, analogous to Fig.~\ref{fig:boston_experiments} in the main text. In this experiment, the physical qubits were fixed along the same pre-defined line on the device for both methods. In alignment with the experiments on the \texttt{ibm\_boston} device, we use $20$ instances for each system size and $2000$ shots per run. The data was acquired without any error mitigation or post-processing.

\section*{Acknowledgments}
We acknowledge the use of IBM Quantum Credits via the IBM Quantum Startups Program for this work. The views expressed are those of the authors and do not reflect the official policy or position of IBM or the IBM Quantum Platform team.
In particular we thank Joachim Schäfer and Junye Huang, for providing technical support and the foundational administration to enable this work. Moreover, we thank Rigetti Computing for providing access to their QPU and in particular Yuvraj Mohan and Kyle Strand for technical and administrative support. This research was funded in part by the Austrian Science Fund (FWF) under Grant-DOI 10.55776/F71. For the purpose
of open access, the authors have applied a CC BY public copyright license to any Author Accepted Manuscript version arising from this submission.

\IEEEtriggeratref{30}
\bibliographystyle{IEEEtran}
\bibliography{references}

\end{document}